\documentclass[useAMS,usenatbib]{mn2e}

\newcommand{\lya}{Ly$\alpha$}
\newcommand{\lyman}{Lyman-$\alpha$}
\newcommand{\msun}{\mbox{M}_{\odot}}
\newcommand{\be}{\begin{enumerate}}
\newcommand{\ee}{\end{enumerate}}
\newcommand{\kms}{\,km\,s$^{-1}$}
\newcommand{\rhot}{$T$-$\rho$}
\newcommand\ionm[2]{#1$\;${\small\rmfamily{#2}}\relax}%

\newcommand{\apj}{ApJ}
\newcommand{\apjl}{ApJL}
\newcommand{\apjs}{ApJS}
\newcommand{\aj}{AJ}

\newcommand{\mnras}{MNRAS}
\newcommand{\nat}{Nature}

\newcommand{\aap}{A\&A}

\usepackage{graphicx}
\usepackage{apjfonts}
\usepackage{amsmath}
\usepackage{amssymb}
\usepackage{verbatim}
\usepackage{times}
\usepackage[usenames]{color}
\usepackage{lscape}

\begin{document}

\title[Scale of the Transverse \lyman\ Forest]{Pressure Support vs.\ Thermal Broadening in the Lyman-$\alpha$ Forest II:\\
Effects of the Equation of State on Transverse Structure}
\author[Peeples et al.]{Molly S.\ Peeples$^{1}$\thanks{E-mail:
    molly@astronomy.ohio-state.edu},  David H.\ Weinberg$^{1}$, Romeel
  Dav\'e$^{2}$, Mark A.\ Fardal$^{3}$, Neal Katz$^{3}$\\
$^{1}$Department of Astronomy and the Center for Cosmology and
  Astro-Particle Physics, The Ohio State University, Columbus,~OH~43210\\
$^{2}$University of Arizona, Steward Observatory, Tucson,~AZ~85721\\
$^{3}$Department of Astronomy, University of Massachusetts, Amherst,~MA~01003}

\pagerange{\pageref{firstpage}--\pageref{lastpage}} \pubyear{2009}

\date{\today}

\maketitle

\label{firstpage}

\begin{abstract}
We examine the impact of gas pressure on the transverse coherence of
high-redshift ($2\leq z \leq 4$) \lyman\ forest absorption along
neighboring lines of sight that probe the gas Jeans scale (projected
separation $\Delta r_p\leq 500 h^{-1}$\,kpc comoving; angular separation
$\Delta\theta\lesssim 30\arcsec$).  We compare predictions from two
smoothed particle hydrodynamics (SPH) simulations that have different
photoionization heating rates and thus different temperature-density
relations in the intergalactic medium (IGM). We also compare spectra
computed from the gas distributions to those computed from the
pressureless dark matter.  The coherence along neighboring sightlines is
markedly higher for the hotter, higher pressure simulation, and lower
for the dark matter spectra.  We quantify this coherence using the flux
cross-correlation function and the conditional distribution of flux
decrements as a function of transverse and line-of-sight (velocity)
separation.  Sightlines separated by $\Delta\theta\lesssim 15\arcsec$
are ideal for probing this transverse coherence.  Higher pressure
decreases the redshift-space anisotropy of the flux correlation
function, while higher thermal broadening increases the anisotropy.  In
contrast to the longitudinal (line-of-sight) structure of the \lya\
forest, the transverse structure on these scales is dominated by
pressure effects rather than thermal broadening.  With the rapid recent
growth in the number of known close quasar pairs, paired line-of-sight
observations offer a promising new route to probe the IGM
temperature-density relation and test the unexpectedly high temperatures
that have been inferred from single sightline analyses.
\end{abstract}


\section{Introduction}\label{sec:intro}
The $1.5\lesssim z\lesssim 6$ intergalactic medium (IGM) is most
commonly studied via the \lyman\ forest, which arises from \lya\
absorption of neutral hydrogen along the line of sight to some distant
source (e.g., a quasar).  Most of the information we have on the nature
of the IGM is from sightlines piercing physically distinct regions of
the IGM.  Absorption features have finite widths, but from individual
sightlines it is difficult to separate the contribution of bulk
velocities (Hubble flow and peculiar velocities) from those of thermal
broadening.  Close quasar pairs can break this degeneracy by probing the
transverse structure of the IGM.  In the mid-1990s, several studies of
quasar groups and lensed quasars definitively showed that the absorbing
structures are coherent over hundreds of kpc
\citep{bechtold94,dinshaw94,fang96,charlton97,crotts98,dodorico98}.  These
observations provided critical support for the physical picture of the
\lya\ forest then emerging from cosmological simulations
\citep{cen94,zhang95,hernquist96} and associated analytic descriptions
\citep{rauch95,reisenegger95,bi97,hui97a}, in which most \lya\
absorption arises in a continuously fluctuating medium of low density
gas rather than in a system of discrete clouds.  More recently,
observations of pairs have been suggested as ways of investigating
measuring the cosmological constant via the \citet{alcock79} effect
\citep{mcdonald99}, the matter power spectrum on small scales
\citep{viel02a}, and the IGM temperature-density relation.  It is on the
last of these that we focus in this paper.

Theoretical models predict that the low density IGM should have a
power-law ``equation of state,''
\begin{equation}\label{eqn:rhot}
T=T_0(1+\delta)^{\alpha},
\end{equation}
 with denser gas being hotter than less dense gas \citep[$\alpha >
0$,][]{katz96,miralda96,hui97b}.  Although this temperature-density
(\rhot) relation is difficult to measure, multiple observations suggest
that it has a higher normalization and a shallower slope than that
expected using the most straightforward assumptions about
photoionization heating \citep{schaye00,mcdonald01,theuns02,bolton08}.
Because the degree of small-scale transverse coherence is set by the
Jeans length, and the Jeans length depends on the temperature of the
gas, studying the transverse structure of the \lya\ forest might give
insight into the IGM \rhot\ relation (J.\ Hennawi, private
communication, 2007).  In Peeples et al.\ (2009, hereafter Paper~I), we
show that while thermal broadening and pressure support both affect the
longitudinal structure of the \lya\ forest, thermal broadening
dominates.  In this paper we investigate the effects of the
temperature-density relation via pressure support and thermal broadening
on the transverse small-scale structure of the \lya\ forest.

The gas temperature affects the Jeans length $\lambda_J$ (and the comoving
Jeans length $\lambda_{J,{\rm comv}}$) via
\begin{eqnarray}
\lambda_J &\equiv& c_s\sqrt{\frac{\pi}{G\rho}}\label{eqn:jeans}\\ 
\Rightarrow\lambda_{J,{\rm comv}} & = & (1+z)\sigma_{\rm th}H_0^{-1}\sqrt{\frac{5\pi}{3}}
\left[\frac{3}{8\pi}\Omega_{m,0}(1+z)^3(1+\delta)\right]^{-1/2}\nonumber\\
 &= & 782\,h^{-1}\,\mbox{kpc}\label{eqn:jeanscomove}\\
 && \times\left(\frac{\sigma_{\rm
 th}}{11.8\,\rm{km}\,\rm{s}^{-1}}\right)\left[\left(\frac{\Omega_{m,0}(1+\delta)}{0.25
\times(1+0)}\right)\left(\frac{1+z}{1+3}\right)\right]^{-1/2}, \nonumber
\end{eqnarray}
where $1+\delta\equiv\rho_{\rm gas}/\bar{\rho}_b$ is the gas overdensity and 
$c_s=\sqrt{[5kT]/[3m]}=\sigma_{\rm th}\sqrt{5/3}$ is the speed of
sound in an ideal gas expressed as a multiple of the 1-D thermal velocity
$\sigma_{\rm th}$, which we have normalized to correspond to $10^4$\,K
\citep{miralda96,schaye01,desjacques05}.  While thermal broadening
affects the observed IGM by smoothing the \lya\ forest in one-dimension
(namely, along the line of sight), pressure support smooths the physical
gas distribution in all three dimensions.  Therefore, while we found in
Paper~I that $\sigma_{\rm th}$ dominates the longitudinal \lya\ forest
structure, we expect $\lambda_J$ to dominate the transverse
structure.  Our simulations indicate that the ``effective'' Jeans length
in the \lya\ forest is smaller than that given in
Equation~(\ref{eqn:jeanscomove}) by a factor of a few, probably owing to
a combination of geometric factors, the universe expanding on the
same timescale as the gas evolves, and the contribution of dark matter to the
gravitational forces \citep[see also][]{gnedin98}.

For $\Omega_m=0.25$ and $\Omega_{\Lambda}=0.75$, the relation between
angular separation $\Delta\theta$ and comoving transverse separation $R$
at $z=2$--4 is approximately
\begin{equation}\label{eqn:dtheta}
\Delta\theta\approx 4.4\arcsec\left(\frac{1+z}{4}\right)^{0.6}\times\left(\frac{R}{100h^{-1}\,\mbox{kpc}}\right).
\end{equation}
Lines of sight with angular separations of 3--10\arcsec\ are needed to
probe the Jeans scale of the IGM.  While this scale is just larger than
the cutoff for the typical Einstein radius of galaxy lenses
\citep*{schneider06}, new searches for binary quasars are revealing
samples of a few to dozens with $\Delta\theta \lesssim 10\arcsec$
\citep{hennawi06,hennawi09}.

This paper is organized as follows.  In \S\,\ref{sec:sims}, we describe
the SPH simulations used, as well as the artificial temperature-density
relations we impose on the gas to isolate the effects of pressure
support and thermal broadening.  In \S\,\ref{sec:results} we discuss how
the temperature-density relation affects the transverse coherence of the
\lya\ forest, with particular focus on flux cross-correlation functions
and conditional flux probability distributions.  We find that, as
expected, the transverse coherence of the \lya\ forest across closely
paired sightlines is dominated by the amount of pressure support in the
absorbing gas.  These conclusions are summarized in \S\,\ref{sec:conc}.
In an Appendix and associated electronic tables, we provide \lya\ forest
spectra extracted from our simulations at several transverse separations
that can be used to create predictions tailored to specific
observational analyses.  We note that Paper~I includes an extensive
discussion of the physical structure of the \lya\ forest in these
simulations, so in this paper we will focus only on those issues
relevant to quasar pair observations.  All distances are given in
comoving coordinates unless otherwise stated.

\section{Simulations}\label{sec:sims}
\begin{figure}
\includegraphics[width=0.48\textwidth]{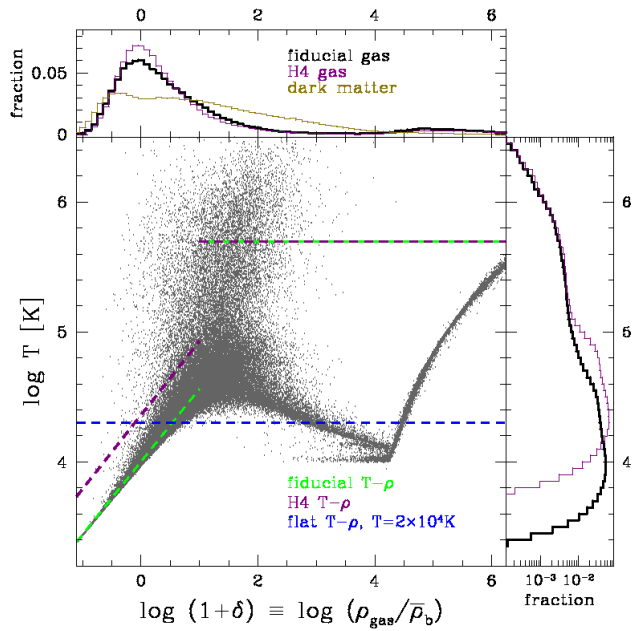}
\caption{\label{fig:rhot}Distribution of particles in the
  temperature-overdensity plane for the fiducial simulation at $z=2.4$,
  with three imposed temperature-density relations over-plotted as
  labelled, as well as the temperature and overdensity distributions for the
  fiducial and H4 simulations.  The hotter H4 gas is preferentially less
  dense than the lower pressure fiducial gas and pressureless dark
  matter.}
\end{figure}

\begin{figure*}
\includegraphics[width=\textwidth]{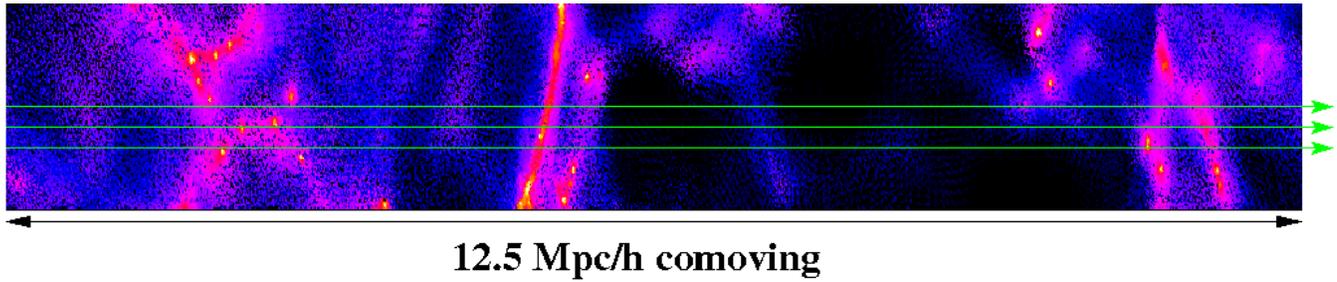}
\caption{\label{fig:slice} A $12.5\times 1\times 1\,h^{-1}$\,Mpc
  comoving section of the fiducial simulation, in \ionm{H}{I} density, with $-8
  < \log n_{\mbox{\scriptsize{HI}}} < 0$ (note that the aspect ratio has
  not been preserved).  The three green sightlines
  are separated by 100\,$h\,^{-1}$\,kpc comoving; at $z=2.4$,
  corresponding to a projected separation of $4.9\arcsec$.}
\end{figure*}

\begin{figure*}
\includegraphics[angle=270,width=\textwidth]{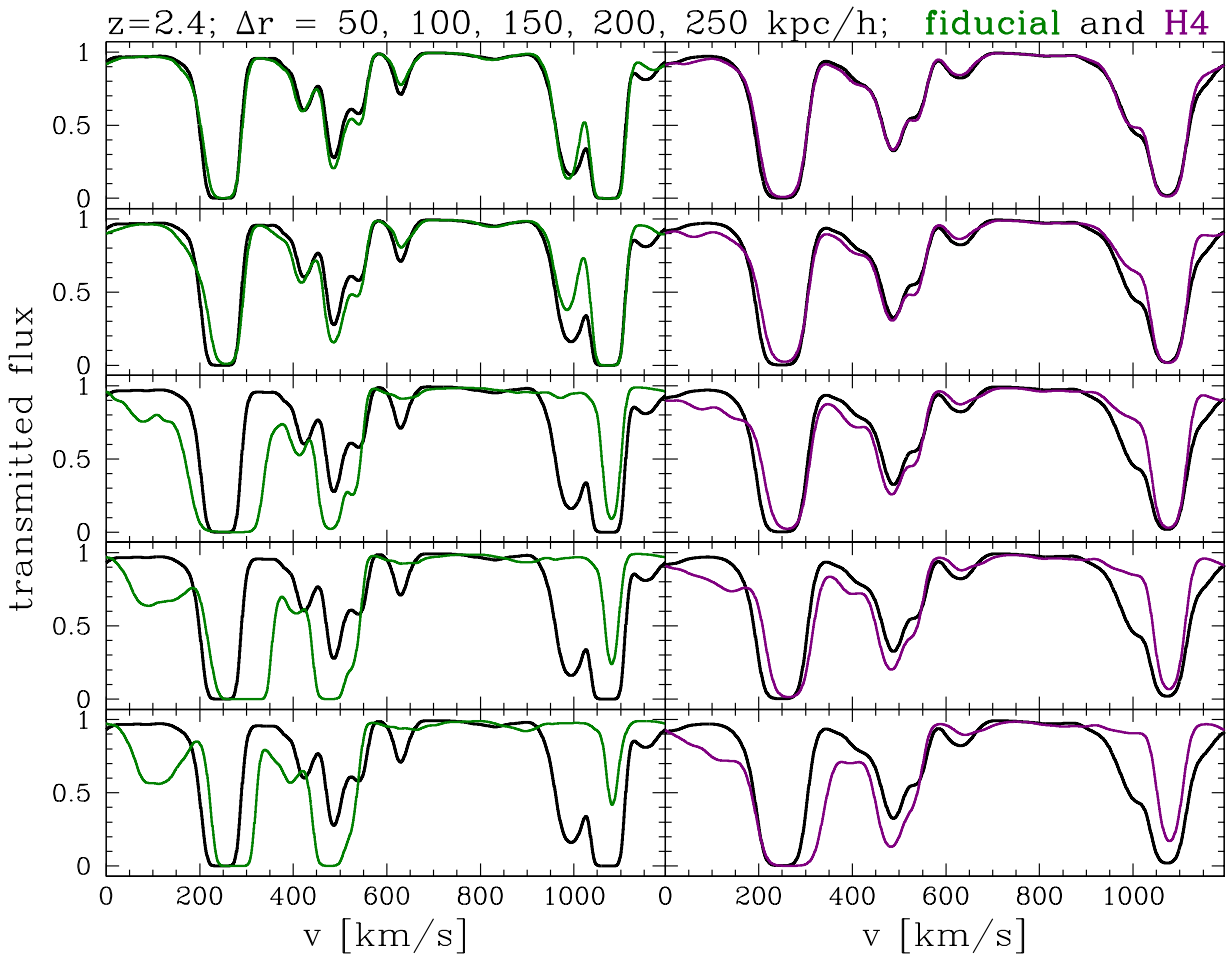}
\caption{\label{fig:spectra} Sample paired lines of sight at $z=2.4$
  separated by $\Delta r = 50$, 100, 150, 200, and 250\,$h^{-1}$\,kpc
  comoving, from top to bottom, with the fiducial simulation on the left
  and the H4 simulation on the right.  Within each column, the black
  spectra are the same ($r = 0$). The top, middle, and bottom green
  spectra correspond to the three green sight lines in
  Fig.~\ref{fig:slice}, with $v=0$ corresponding to the left-hand side
  of Fig.~\ref{fig:slice}. }
\end{figure*}

We use the same $2\times288^3$ particle $12.5\,h^{-1}$\,Mpc comoving
smoothed particle hydrodynamic (SPH) simulations evolved with {\sc
Gadget-2} \citep{springel05} as in Paper~I; hence we only present a
basic description here. Throughout we adopt a $\Lambda$CDM cosmology of
$(\Omega_m,\Omega_{\Lambda},\Omega_b,h,\sigma_8,n_s) =
(0.25,0.75,0.044,0.7,0.8,0.95)$, which is in good agreement with the
{\em Wilkinson Microwave Anisotropy Probe} 5-year results
\citep{hinshaw09}.  This cosmology leads to a gas particle mass of
$1.426\times10^6\,\msun$, which is much less than the expected typical
Jeans mass of $\sim7\times10^9\,\msun$.  As a convergence test, we use a
$2\times144^3$ particle simulation that is otherwise identical to our
fiducial simulation.  The distribution of particles in the
temperature-density plane for our ``fiducial'' simulation at $z=3$ is
shown in Figure~\ref{fig:rhot}.  The ``H4'' simulation has the same
initial conditions as the fiducial one, but the heating rate from
photoionization by the UV background is four times higher than in the
fiducial simulation.  An obvious consequence of this higher heating rate
is that the H4 gas has higher temperatures than the fiducial gas.  A
more subtle effect, also shown in Figure~\ref{fig:rhot}, is that the
hotter gas has a larger Jeans length and is hence smoother, with a
smaller fraction of the gas at high overdensity.  We therefore adopt
three artificial temperature-density relations to isolate the effects of
pressure support, thermal broadening, and the underlying overdensity
distribution.  As in Paper~I, the fiducial and H4 temperature-density
relations mimic the ones found in those simulations, while the flat
$T=2\times10^4$\,K relation is used so that we can study the effects of
thermally-broadened pressure support in the absence of
overdensity-dependent thermal broadening.  We impose these
temperature-density relations on the fiducial and H4 gas distributions,
as well as the fiducial dark matter, by assigning temperatures based
solely on the local gas (or dark matter) overdensity, as demonstrated in
Figure~\ref{fig:rhot}.  For the fiducial and H4 relations, we set all
gas with $1+\delta\ge 10$ to a ``shocked'' temperature $T=5\times
10^5$\,K, while for the flat relation we set all the gas to $T=2\times
10^4$\,K, regardless of density.

\begin{table*}
 \centering
 \caption{\label{tab:rhot}Observed mean flux decrements $\langle D \rangle\equiv
    \langle 1-e^{-\tau}\rangle$ at $z=2.4$,\,3, and 4 are from
    \citet{mcdonald00}; the $z=2$ measurement is from \citet{faucher08}
    without correcting for metal absorption.  The observed
    temperature-density relations ($T = T_0[1+\delta]^{\alpha}$) at
    $z=2.4$,\,3, and 4 are from \citet{mcdonald01}, with the $z\sim 2$
    measurement from \citet{ricotti00}.}
 \begin{tabular}{ccrcccccc}
\hline\hline 
$z$ & $\langle D\rangle$ & observed $T_0$ [K] & observed $\alpha$ & fiducial
$T_0$ [K] & fiducial $\alpha$ & H4 $T_0$ [K] & H4 $\alpha$ & $\Delta\theta$ at \\
 & & &  & & & & & $100\,h^{-1}$\,kpc\\\hline
4.0 & $0.525\pm 0.012$ &   $17400\pm 3900$ & $0.43\pm 0.45$ & $11700$ & $0.54$ & 28200 & 0.55 &  3.9\arcsec\\\hline
3.0 & $0.316\pm 0.023$ &  $18300\pm 1800$ & $0.33\pm 0.26$ & $11000$ & $0.57$ & 25000 & 0.57 & 4.4\arcsec\\
    &  & or $18400\pm 2100$ & $0.29\pm 0.30$ & & & & & \\\hline
2.4 & $0.182\pm 0.021$ &    $17400\pm 1900$ & $0.52\pm 0.14$ & $10000$ & $0.56$ & 23000 & 0.57 &4.9\arcsec\\
    &  & or $19200\pm 2000$ & $0.51\pm 0.14$ & & & &  &\\\hline
2.0 & $0.144\pm 0.024$ & $\quad 17700\quad\quad\quad$\hfill & $0.32\pm 0.30$ & $8913$ & $0.56$ & $21380$ & 0.57 & 5.4\arcsec\\\hline
\end{tabular}
\end{table*}

At each redshift---$z=2$, 2.4, 3, and 4---we consider 200 lines of sight
with paired sightlines separated by $\Delta r = 50$, 100, 125, 150, 175,
200, 250, 300, 400, and 500\,$h^{-1}$\,kpc comoving for a total of 2200
sightlines per redshift for each of the overdensity--\rhot\ combinations
discussed above.  In all cases we adjust the intensity of the UV
background so that the mean flux decrement matches observational
estimates (see Paper~I for details).  In Table~\ref{tab:rhot}, we list
the adopted mean decrements and parameters for the observed, fiducial,
and H4 temperature-density relations, as well as the projected angular
separation at 100\,$h^{-1}$\,kpc comoving.

\section{Structure of the IGM \& Transverse \lya\ Coherence}\label{sec:results}
To gain insight into the effects of temperature on the transverse
coherence of the \lya\ forest, we look at how paired sightlines differ
in the fiducial and H4 simulations in \S\,\ref{sec:spectra}.  We then
quantify these differences by examining the relative changes in the flux
decrement cross-correlation function in \S\,\ref{sec:crosscorr} and the
relative transverse coherence of the conditional flux decrement
probability distribution in \S\,\ref{sec:pdgived}.

\subsection{Spectra}\label{sec:spectra}
Before delving into statistical measures of the transverse \lya\ forest,
it will be instructive to first consider the underlying physical
structures.  In Figure~\ref{fig:slice}, we show a small section of the
fiducial simulation at $z=2.4$, where brighter regions correspond to higher
\ionm{H}{I} densities.  The observed transmitted flux is calculated as
simply
\begin{equation}\label{eqn:tau}
F = e^{-\tau_{{\rm Ly}\alpha}},
\end{equation}
where $\tau_{{\rm Ly}\alpha}$ is the optical depth to \lya\ photons.
As discussed in detail in Paper~I, 
\begin{eqnarray}\label{eqn:tauGP}
\tau_{{\rm Ly}\alpha} & \propto & n_{\mbox{\tiny HI}}\\
&\propto & \left(\frac{T_0}{10^4\,\mbox{K}}\right)^{-0.7} \left(1+\delta\right)^{2-0.7\alpha}.\nonumber
\end{eqnarray}
The three sightlines in Figure~\ref{fig:slice} give rise to the top,
middle, and bottom green spectra at $z=2.4$ in the left-hand panel of
Figure~\ref{fig:spectra}.  Figure~\ref{fig:spectra} shows how pairs of
spectra become more dissimilar as their transverse separation, $\Delta r$,
increases, and how this dissimilarity differs between the fiducial and
H4 simulations.  Spectral features remaining coherent over large scales
correspond to physical structures that are parallel to the plane of the
sky (see, e.g., the structures at $400<v<600$\,km\,s$^{-1}$), while
spectral features disappearing from one sightline to the next correspond
to physical structures that are more parallel to the line of sight (see,
e.g., the features at $v < 300$\,km\,s$^{-1}$).  In general, the H4
spectra remain more similar than the fiducial ones as $\Delta r$
increases; we aim to determine to what extent this relatively higher
coherence owes to pressure support rather than to the fact that the H4
spectra are individually inherently smoother because they have more
thermal broadening than the fiducial spectra.

\subsection{Cross-Correlation Functions}\label{sec:crosscorr}
\begin{figure}
\includegraphics[angle=270,width=0.48\textwidth]{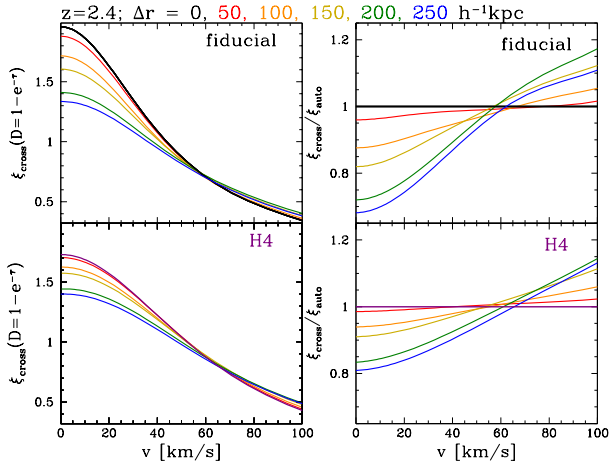}
\caption{\label{fig:crosscorrcomp}The cross-correlation function
  $\xi_{\rm cross} \equiv\langle D_1(v)D_2(v+\Delta v)\rangle/\langle
  D\rangle^2$ ({\em left}) and $\xi_{\rm cross}/\xi_{\rm auto}$ ({\em
  right}) for the fiducial ({\em top}) and H4 ({\em bottom}) simulations
  for six different transverse separations $\Delta r = 50$, 100, 150,
  200, and 250\,$h^{-1}$\,kpc comoving.  Note that the vertical scale in
  the left panels does not extend to 0.}
\end{figure}

A common method for studying the transverse structure of the IGM is to
look at the flux decrement cross-correlation function,
\begin{equation}\label{eqn:crosscorr}
\xi_{\rm cross} \equiv \frac{\langle D_1(v)D_2(v+\Delta v)\rangle}{\langle D\rangle^2},
\end{equation}
where $D\equiv 1-F = 1-\exp(-\tau)$ and the two sightlines are separated
by some $\Delta r$ \citep{miralda96,rollinde03}. At $\Delta r = 0$,
$\xi_{\rm cross}$ is just the auto-correlation function.  The
cross-correlation functions for the $z=2.4$ fiducial and H4 simulations
at a range of $\Delta r$ are shown in the left-hand panels of
Figure~\ref{fig:crosscorrcomp}.  At small $\Delta r$ and $\Delta v$, the
gas in the fiducial simulation has a higher $\xi_{\rm cross}$ than the
gas in the H4 simulation because the smoother gas distribution has less
rms density fluctuation.  At larger $\Delta r$ and $\Delta v$, the H4
gas has a higher $\xi_{\rm cross}$ owing to its greater coherence, as is
visually evident in Figure~\ref{fig:spectra}.  To quantify this relative
change, $\xi_{\rm cross}/\xi_{\rm auto}$ for the same selection of
$\Delta r$ is plotted in the right-hand panels of
Figure~\ref{fig:crosscorrcomp}. Although in real space $\xi_{\rm
cross}(\Delta r) = \xi_{\rm auto}(H\Delta r)$, in velocity space
redshift distortions preferentially suppress the auto-correlation
function relative to the cross-correlation function, causing $\xi_{\rm
cross} > \xi_{\rm auto}$ for some regions of parameter space
\citep{mcdonald99,marble08b}.  The hotter, higher pressure, H4
simulation has a higher relative coherence (larger $\xi_{\rm
cross}/\xi_{\rm auto}$) at small $\Delta v$ than the colder fiducial
simulation.

\begin{figure}
\includegraphics[width=0.48\textwidth]{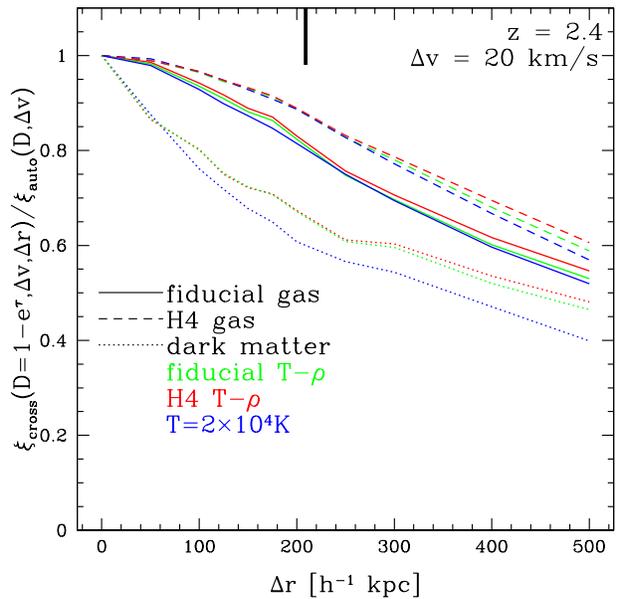}
\caption{\label{fig:crosscorrdrnormz24}Normalized cross-correlation
  functions, $\xi_{\rm cross}/\xi_{\rm auto}$, as a function of the
  transverse separation $\Delta r$ for $\Delta v = 20$\,km\,s$^{-1}$ at
  $z=2.4$.  The line type (solid, dashed, or dotted) denotes the adopted
  overdensity field and the line color the adopted \rhot\ relation (see
  \S\,\ref{sec:sims} for details).  The thick black tickmark denotes
  $\Delta r=H\Delta v$.}
\end{figure}

We compare a wider range of models in
Figure~\ref{fig:crosscorrdrnormz24}, where we plot $\xi_{\rm
cross}/\xi_{\rm auto}$ as a function of $\Delta r$ at $\Delta v =
20$\kms\ at $z=2.4$; this corresponds to taking a slice at $\Delta
v=20$\kms\ in the right-hand panels of Figure~\ref{fig:crosscorrcomp}.
To elucidate whether the H4 gas is more strongly correlated because it
has higher pressure, or because, as shown in detail in Paper~I, hotter
gas has more thermal broadening and therefore less small-scale
structure, we also look at a range of \rhot\ relations and overdensity
fields.  In this and in several of the following figures the line type
(solid, dashed, or dotted) corresponds to the adopted overdensity field,
either the gas overdensities from the fiducial simulation (solid lines),
the gas overdensities from the H4 simulation (dashed lines) or the dark
matter overdensities from the fiducial simulations.  The line color
corresponds to the adopted equation of state, i.e. the \rhot\ relation,
either the artificial fit to the relation from the fiducial simulation
(green lines), the fit to the relation from the H4 simulation (red
lines), or assuming a constant temperature of $T=2\times10^4$\,K (blue
lines).

\begin{figure}
\includegraphics[width=0.48\textwidth]{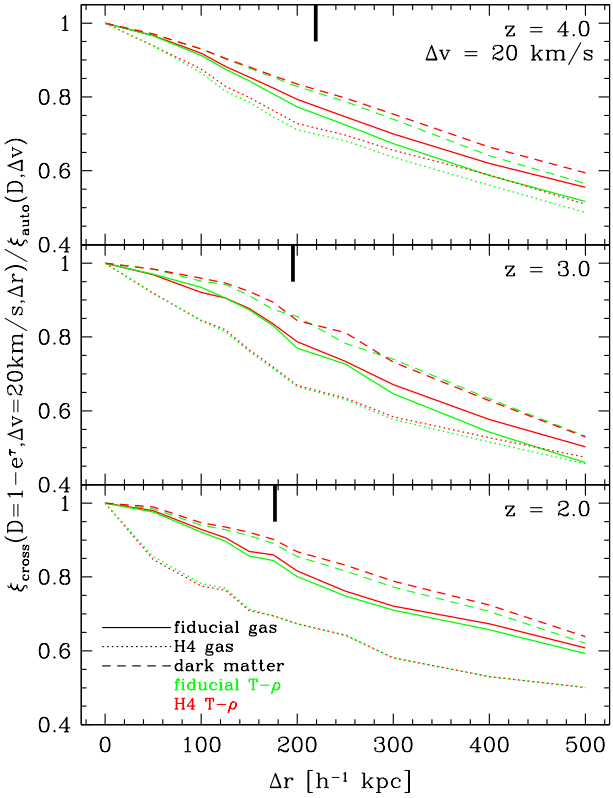}
\caption{\label{fig:crosscorrdrnormz234}Normalized cross-correlation
  functions, $\xi_{\rm cross}/\xi_{\rm auto}$, as a function of the
  transverse separation $\Delta r$ for $\Delta v = 20$\,km\,s$^{-1}$ at
  $z=4$, 3, and 2, from top to bottom.  The models are denoted as in
  Fig.~\ref{fig:crosscorrdrnormz24}.  The thick black tickmark denotes
  $\Delta r=H\Delta v$.}
\end{figure}

In Figure~\ref{fig:crosscorrdrnormz24}, the three choices for the
overdensity field clearly separate into three distinct groups, with the
highest-pressure H4 gas having the most transverse coherence and the
zero-pressure dark matter having the least.  Even non-thermally
broadened spectra (which we have not plotted to avoid visual confusion)
show the same relative decrease in coherence with increase in transverse
separation as other spectra with the same underlying gas distributions.
Within each overdensity group, $\xi_{\rm cross}/\xi_{\rm auto}$
increases with increasing thermal broadening (e.g., the imposed H4
\rhot\ relation yields higher $\xi_{\rm cross}/\xi_{\rm auto}$ at all
$\Delta r$ than imposing the fiducial \rhot).  However, lowering the
resolution leads to offsets from the fiducial case by about the same
amount as imposing different \rhot\ relations.  The same trends are seen
at $z=2$,\,3, and 4, as shown in Figure~\ref{fig:crosscorrdrnormz234}.
In general, we find this delineation is clearer at $10\lesssim\Delta v
\lesssim 30$\kms\ than at $\Delta v=0$ or at larger velocity
separations.  The stark separation by overdensity distribution of the
normalized cross-correlation function as a function of transverse
separation is a clear sign that pressure plays an important role in the
transverse structure of the \lya\ forest.

\begin{figure}
\includegraphics[width=0.48\textwidth]{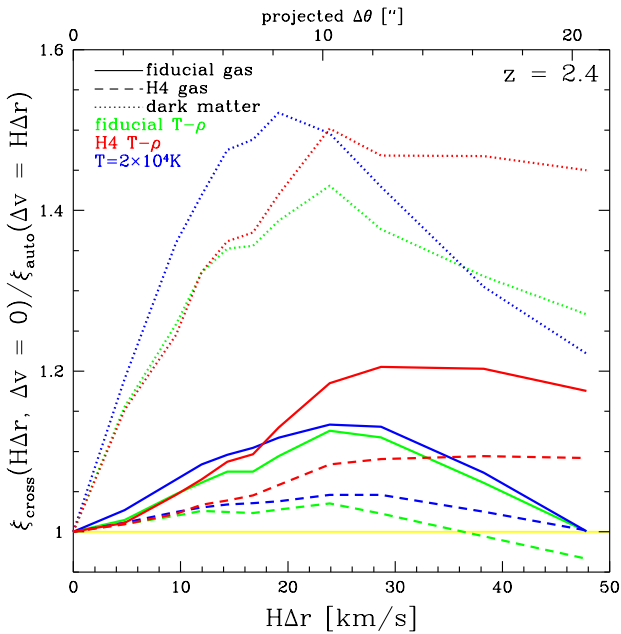}
\caption{\label{fig:corranisoz24}Normalized cross-correlation functions
  showing the anisotropy of the \lya\ forest, $\xi_{\rm cross}/\xi_{\rm
  auto}$, as a function of $\Delta v = H\Delta r$, at $z=2.4$. The models are
  denoted as in Fig.~\ref{fig:crosscorrdrnormz24}}
\end{figure}

\begin{figure}
\includegraphics[angle=270,width=0.48\textwidth]{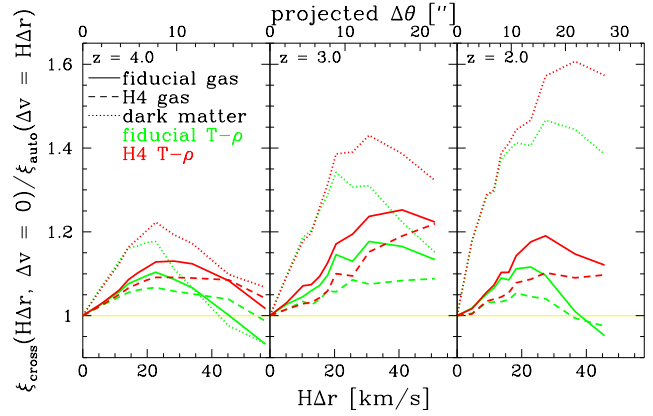}
\caption{\label{fig:corranisoz234}Normalized cross-correlation functions
  showing the anisotropy of the \lya\ forest, $\xi_{\rm cross}/\xi_{\rm
  auto}$, as a function of $\Delta v = H\Delta r$, at $z=4$, 3, and 2.
  The models are denoted as in Fig.~\ref{fig:crosscorrdrnormz24}}
\end{figure}

A common use for the flux decrement cross-correlation function is to
measure the anisotropy in the \lya\ forest caused by line-of-sight
velocity distortions \citep{coppolani06,dodorico06}, such as for the
\citet{alcock79} test \citep{hui99,mcdonald99}.  In
Figure~\ref{fig:corranisoz24}, we compare $\xi_{\rm cross}(\Delta
v=0,\Delta r)$ to the autocorrelation function at the same scale,
$\xi_{\rm auto}(H\Delta r)$, at $z=2.4$; higher values indicate higher
levels of anisotropy.  As in Figure~\ref{fig:crosscorrdrnormz24}, the
models separate into groups of overdensity, and the differences between
our low-resolution and fiducial cases are comparable with imposing
different \rhot\ relations on the fiducial
gas. Figure~\ref{fig:corranisoz234} shows the same statistic for a
smaller set of models at $z=2$, 3, and 4.  As mentioned above, in real
space $\xi_{\rm cross}(\Delta r)=\xi_{\rm auto}(H\Delta r)$, but in
(observed) velocity space, redshift distortions introduce anisotropy
\citep{marble08b}.  Because thermal broadening is an inherently
one-dimensional anisotropic phenomenon, higher thermal broadening leads
to higher anisotropy; we see this effect for each overdensity
distribution in Figures~\ref{fig:corranisoz24} and
\ref{fig:corranisoz234}. Pressure, on the other hand, is a
three-dimensional inherently isotropic phenomenon: higher pressure
therefore leads to less anisotropy, as we see from comparing the results
for the dark matter, fiducial, and H4 overdensity fields. For $H\Delta r
\lesssim 20$\kms\ ($\Delta\theta\lesssim 10\arcsec$), the impact of
thermal broadening on anisotropy is generally smaller than the impact of
pressure.

\vspace*{1cm}

\subsection{Conditional Flux Probability Distributions}\label{sec:pdgived}
\begin{figure*}
\begin{center}
\includegraphics[width=0.95\textwidth]{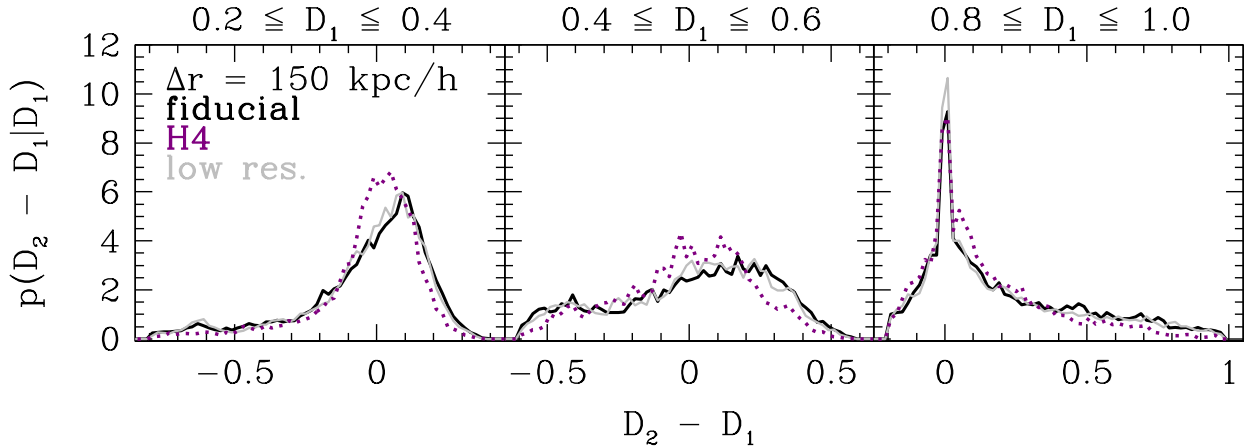}
\end{center}
\caption{\label{fig:pdgivedvpd}Flux decrement difference probability
  distributions, $p(D_2 - D_1)$ v.\ $(D_2 - D_1)$, at $z=2.4$, for
  $\Delta v = 0$ and $\Delta r=150$.  The three columns show
  $D_1\in[0.2,0.4)$ ({\em left}), $D_1\in[0.4,0.6)$ ({\em middle}), and
  $D_1\in[0.8,1.0]$ ({\em right}), with fiducial in black, H4 in purple
  ({\em dotted}), and the low resolution simulation in grey.  }
\end{figure*}

The structure of one-point flux probability distribution function (PDF)
depends on both the thermal history and current thermal state of the
gas, leading to a complex relationship between the effects of pressure
support and thermal broadening on the PDF (Paper~I).  On the other hand,
the interpretation of the conditional probability distribution function
between paired sightlines \citep{miralda97} is relatively
straightforward.  For example, if for strongly absorbed pixels with $0.8
\leq D_1 < 1.0$, the pixels separated by $\Delta v$ on a sightline
$\Delta r$ away are more strongly absorbed than randomly expected, then
this might be a signature of strong transverse coherence and thus a
large Jeans length.  In Figure~\ref{fig:pdgivedvpd} we plot the flux
decrement difference probability distributions, $p(D_2 - D_1)$, for
$\Delta v = 0$ and $\Delta r=150\,h^{-1}$\,kpc and several bins of
$D_1$, for the fiducial, H4, and low resolution gas at $z=2.4$. By
looking at the PDF of the decrement differences, we can easily quantify
the similarity of flux decrement pairs.  The more strongly the
distribution peaks around $D_2=D_1$, the more coherent are the
transverse structures.  While the differences between the two
simulations are not dramatic, the H4 model is consistently more strongly
peaked around $D_2=D_1$, with a stronger signature at low $D_1$.  For
most choices of $D_1$, this difference is much more pronounced than the
difference between the fiducial and low-resolution simulations,
indicating that our $288^3$ particle simulations give robust results for
this statistic.

\begin{figure*}
\begin{center}
\includegraphics[height=0.9\textheight]{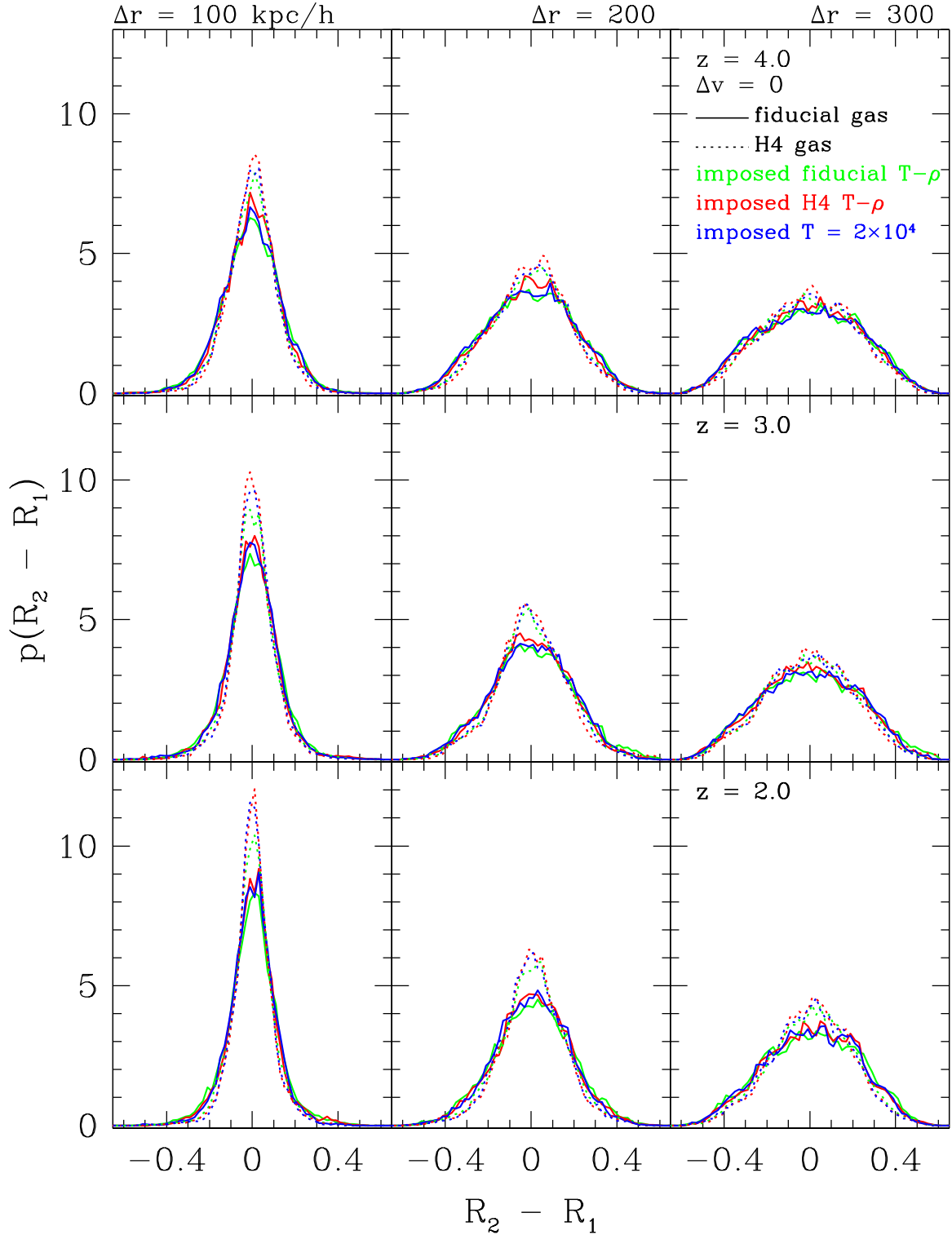}
\end{center}
\caption{\label{fig:pdgived_diff_rank_zoom}Probability distributions for
  differences in flux decrement rank, $p(R_2 - R_1)$ vs.\ $(R_2 - R_1)$
  for $\Delta v = 0$, $0.4 \leq R_1\leq 0.6$, and $z=4$, 3, and 2.0 (top
  to bottom).  The three columns show comoving transverse separations
  $\Delta r = 100$,\,200, and 300\,$h^{-1}$\,kpc (left to
  right). Regardless of the imposed temperature-density relation
  (indicated by line color), the H4 gas ({\em dotted lines}) is more
  strongly peaked around $R_2 = R_1$ than the fiducial gas ({\em solid
  lines}) because the hotter gas has longer Jeans lengths and thus
  higher transverse coherence.}
\end{figure*}

\begin{figure*}
\begin{center}
\includegraphics[height=0.9\textheight]{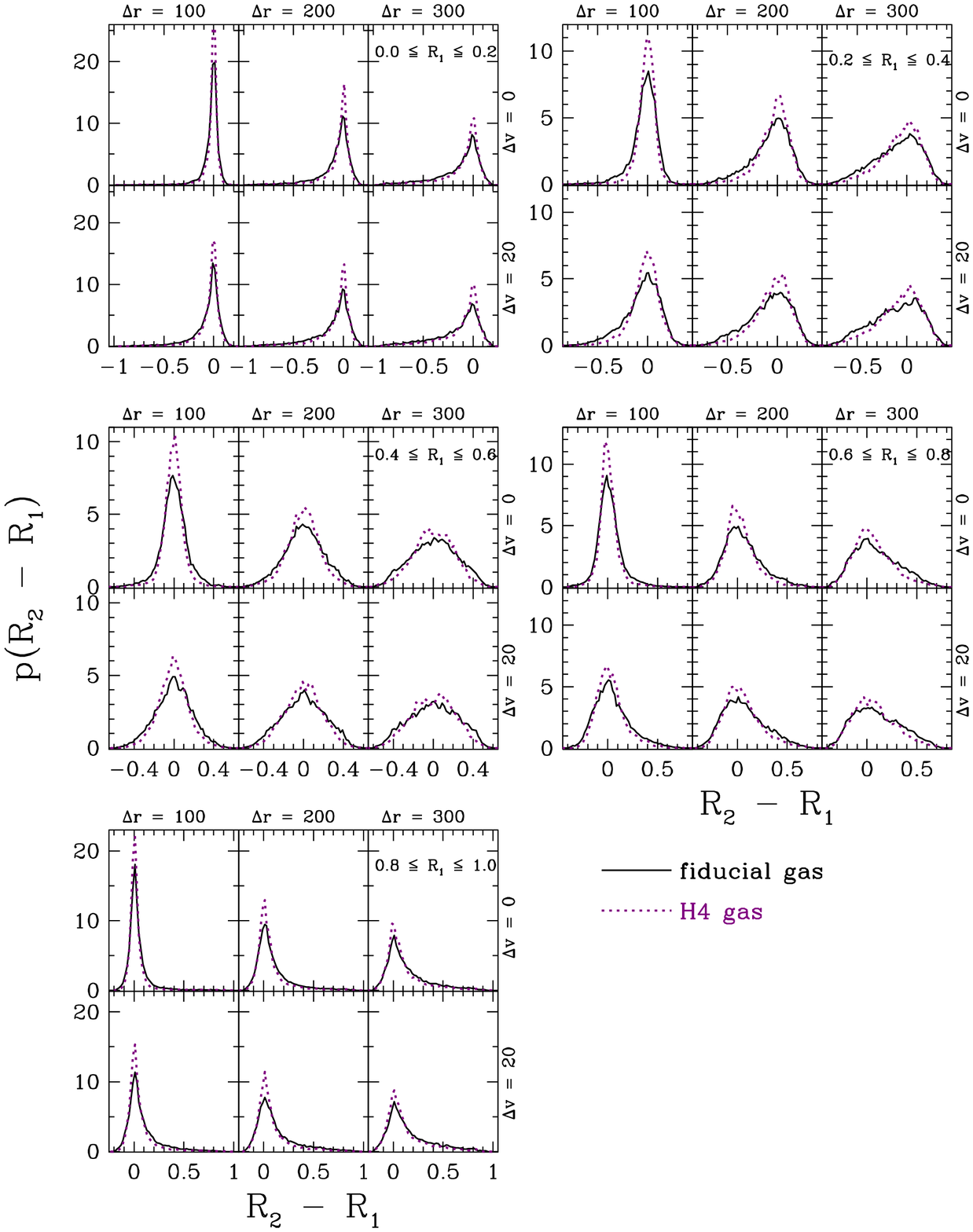}
\end{center}
\caption{\label{fig:pdgived_diff_rank}Probability distributions for
  differences in flux decrement rank, $p(R_2 - R_1)$ v.\ $(R_2 - R_1)$
  for $\Delta r = 100$,\,200, and 300\,$h^{-1}$\,kpc\,comoving and
  $\Delta v = 0$ and 20\kms as labelled, for a range of $R_1$ at $z=2.4$.
  The H4 gas ({\em purple, dotted}) is more strongly peaked around $R_2
  = R_1$ than the fiducial gas ({\em black, solid}) because the hotter
  gas has longer Jeans lengths and thus a higher transverse coherence.}
\end{figure*}

In Paper~I we showed that the flux decrement PDFs for each of these
\rhot\ relations are fairly distinct, so some of the model differences
in $p(D_2-D_1)$ could reflect differences in the underlying flux PDFs.
We can remove this effect by converting from flux decrement $D$ to pixel
rank $R\equiv p(<D)$, the fraction of pixels with a flux decrement lower
than $D$. Of course, a fully successful model should reproduce the
observed PDF, but here we wish to focus on transverse coherence and,
therefore, remove any differences in the PDF caused by different
temperature-density relations.  Figure~\ref{fig:pdgived_diff_rank_zoom}
shows that spectra generated from the H4 gas have rank difference
distributions more strongly peaked around $R_2-R_1=0$ than spectra from
the lower pressure, fiducial gas spectra.  Changing the imposed
temperature-density relation has less effect on the distributions than
changing the underlying gas distribution, implying that (as expected)
pressure rather than thermal broadening accounts for the larger
transverse coherence.  In general, the larger coherence appears at all
redshifts, but it weakens with increasing $\Delta r$.  Non-thermally
broadened spectra (not shown) have broader $R_2-R_1$ distributions, so
thermal broadening does play some role in transverse coherence.  As with
previous statistics, however, the lower resolution spectra differ from
the fiducial case by about as much as spectra generated from different
imposed \rhot\ relations.  Figure~\ref{fig:pdgived_diff_rank} presents
the $z=2.4$ predictions in greater detail for the fiducial and H4 gas
only, plotting five ranges of $R_1$ and comparing $\Delta v=0$ to
$\Delta v\sim 20$\,km\,s$^{-1}$.  The models are most easily
distinguished at small $\Delta v$ and at small $\Delta r$.

In general, the slope of the temperature-density relation is much more
difficult to observationally constrain than $T_0$ because most methods
for measuring the \rhot\ relation are sensitive to only a limited range
of $\tau_{\mbox{\tiny HI}}$ and hence $1+\delta$.  Because (up to
saturation) we can limit ourselves to a particular range of optical
depth and thus $1+\delta$ when using {\em conditional} rank
distributions, it might be possible to use this technique to constrain
the slope of the temperature-density relation.  We cannot test this
possibility using our current simulations because the fiducial and H4
temperature-density relations have similar slopes at all redshifts (see
Table~\ref{tab:rhot}).

\section{Conclusions}\label{sec:conc}
Recent efficient searches for binary quasars have yielded large samples
of quasars with angular separations of $\lesssim 10$\arcsec\
\citep{hennawi06,hennawi09}.  The closely paired \lyman\ forest
sightlines from such quasar pairs are ideal for studying the small-scale
transverse structure of the intergalactic medium.  We have shown using a
set of smoothed particle hydrodynamic simulations with different
equations of state that the coherence of transverse structure at these
scales is determined primarily by the level of pressure support, i.e.\
the Jeans length, of the absorbing gas and is relatively insensitive to
the amount of thermal broadening along the line of sight.  Given the
surprisingly high temperatures implied by single-sightline analyses (see
discussions in \S\ref{sec:intro} and Paper~I), it would be valuable to
investigate the thermal state of the IGM by this largely independent
method.

Flux correlation functions (measured by, e.g.,
\citealt{rollinde03,coppolani06,dodorico06,marble08a}) and the two-point
distributions of flux decrements or pixel ranks can all be used to
distinguish among gas distributions with different temperatures like the
fiducial and H4 models considered here.  Because the redshift ranges and
pair separations will be dictated by the specifics of the available data
sample, we provide in the form of electronic tables our simulated
spectra along paired lines of sight from the two simulations, which can
be used to generate predictions tailored to a particular data sample.
These samples of spectra are described in \S\ref{sec:sims} and the
caption to Table~\ref{tab:rhot}.  We caution that the finite size of our
simulation volume could cause statistical fluctuations and systematic
effects on our predictions, especially at large $\Delta r$ and $\Delta
v$; we will investigate this point in future work with larger
simulations.  The large increase in known quasar pairs, the ability of
large telescopes to measure \lyman\ absorption spectra of relatively
faint background sources, which have a high surface density on the sky,
and the massive quasar sample expected from the Baryon Oscillation
Spectroscopic Survey \citep{schlegel10} will change the \lya\ forest from a
one-dimensional phenomenon to a three-dimensional phenomenon, opening
new opportunities to constrain cosmology and the physics of the diffuse
intergalactic medium.

\section*{Acknowledgments}
We acknowledge a seminar by Joe Hennawi at the Ohio State Center for
  Cosmology and Astro-Particle Physics, which inspired this
  investigation; we also thank Joe Hennawi and Eduardo Rozo for helpful
  discussions and comments.  We are grateful to the anonymous referee
  for thoughtful suggestions on the text. This work has been supported
  in part by NSF grant AST-0707985 and NASA ADP grant NNX08AJ44G.

\appendix 

\section[]{Format of associated electronic tables}
We provide in the form of electronic tables our simulated spectra, the
\ionm{H}{I}\ optical depth as a function of velocity and transverse
separation, along paired sightlines from the fiducial and H4 simulations
at $z=2$, 2.4, 3, and 4.  We show in Table~\ref{tab:spec} a portion of
the spectra from the fiducial simulation at $z=2$.

\begin{onecolumn}
\begin{landscape}
\label{lastpage}
\begin{table*}
 \centering
 \caption{\label{tab:specfidz2}Sample paired spectra for the fiducial
   simulation at $z=2$. Each set of 1250 lines corresponds to an
   independent sightline and its pairs.  This table is published in
   their entirety online; a portion is shown here for guidance regarding
   its form and content.}
\begin{tabular}{llllllllllll}
\hline\hline &
\multicolumn{11}{c}{$\tau_{\rm HI}$ at transverse separation $\Delta r$ in $h^{-1}$\,kpc comoving}\\
\cline{2-12}\\
$v$ [\kms] & $0$ & $50$ & $100$ & $125$ & $150$ & $175$ & $200$ & $250$ & $300$ & $400$ & $500$\\\hline
$0.456435$ & $0.0331219$ & $0.0285025$ & $0.0288973$ & $0.0297425$ & $0.0303896$ & $0.0308683$ & $0.0316429$ & $0.0371795$ & $0.0469323$ & $0.0521755$ & $0.0474118$\\
$1.3693$ & $0.0314817$ & $0.0274538$ & $0.0280237$ & $0.0289396$ & $0.0296675$ & $0.0302456$ & $0.0311112$ & $0.0367447$ & $0.0462802$ & $0.0511294$ & $0.0459319$\\
$2.28217$ & $0.0301246$ & $0.0265771$ & $0.0272563$ & $0.0282159$ & $0.0290117$ & $0.0296889$ & $0.0306551$ & $0.0364166$ & $0.0457752$ & $0.0502636$ & $0.0446447$\\
$3.19504$ & $0.0290066$ & $0.0258401$ & $0.0265773$ & $0.0275609$ & $0.028416$ & $0.0291931$ & $0.0302679$ & $0.0361835$ & $0.0454074$ & $0.0495666$ & $0.0435458$\\
$4.1079$ & $0.028088$ & $0.0252151$ & $0.0259722$ & $0.0269665$ & $0.0278759$ & $0.0287538$ & $0.0299433$ & $0.0360341$ & $0.0451657$ & $0.049025$ & $0.0426295$\\
$5.02077$ & $0.0273332$ & $0.0246789$ & $0.0254291$ & $0.026426$ & $0.0273876$ & $0.0283671$ & $0.0296758$ & $0.0359581$ & $0.0450378$ & $0.0486238$ & $0.041889$\\
$5.93364$ & $0.0267117$ & $0.0242127$ & $0.0249388$ & $0.0259346$ & $0.026948$ & $0.0280296$ & $0.0294603$ & $0.0359456$ & $0.0450103$ & $0.0483473$ & $0.0413157$\\
$6.8465$ & $0.0261978$ & $0.0238013$ & $0.0244938$ & $0.0254882$ & $0.0265548$ & $0.0277383$ & $0.0292924$ & $0.0359874$ & $0.0450688$ & $0.0481795$ & $0.0408999$\\
$7.75937$ & $0.0257704$ & $0.0234329$ & $0.0240884$ & $0.0250838$ & $0.0262058$ & $0.0274906$ & $0.0291683$ & $0.0360749$ & $0.0451985$ & $0.0481045$ & $0.0406308$\\
$8.67224$ & $0.0254127$ & $0.0230988$ & $0.0237179$ & $0.024719$ & $0.025899$ & $0.0272841$ & $0.0290847$ & $0.0361998$ & $0.0453839$ & $0.0481074$ & $0.0404972$\\
\ldots & \ldots & \ldots & \ldots & \ldots & \ldots & \ldots & \ldots & \ldots & \ldots & \ldots & \ldots \\
$1139.71$ & $0.0374395$ & $0.0312689$ & $0.031053$ & $0.031642$ & $0.0320651$ & $0.0323369$ & $0.032963$ & $0.0384191$ & $0.0487098$ & $0.0548428$ & $0.0509672$ \\
$1140.63$ & $0.0350917$ & $0.0297604$ & $0.0298985$ & $0.0306379$ & $0.0311856$ & $0.0315631$ & $0.0322576$ & $0.0377333$ & $0.0477398$ & $0.0534111$ & $0.0490887$ \\
$0.456435$ & $0.0097211$ & $0.0087474$ & $0.00850357$ & $0.00869964$ & $0.0087597$ & $0.00884541$ & $0.00929591$ & $0.0102907$ & $0.01099$ & $0.0110082$ & $0.00980152$ \\
$1.3693$ & $0.0102372$ & $0.00920984$ & $0.00901555$ & $0.00928603$ & $0.00938722$ & $0.00950945$ & $0.0100469$ & $0.0111618$ & $0.0118516$ & $0.0117359$ & $0.0103104$ \\
\ldots & \ldots & \ldots & \ldots & \ldots & \ldots & \ldots & \ldots & \ldots & \ldots & \ldots & \ldots \\\hline
\end{tabular}
\end{table*}

\begin{table*}
 \centering
 \caption{\label{tab:specH4z2} Same as Table~\ref{tab:specfidz2} but for
 the H4 simulation at $z=2$.}
\begin{tabular}{llllllllllll}
\hline\hline &
\multicolumn{11}{c}{$\tau_{\rm HI}$ at transverse separation $\Delta r$ in $h^{-1}$\,kpc comoving}\\
\cline{2-12}\\
$v$ [\kms] & $0$ & $50$ & $100$ & $125$ & $150$ & $175$ & $200$ & $250$ & $300$ & $400$ & $500$\\\hline
$0.456435$ & $0.0748727$ & $0.0594009$ & $0.0534443$ & $0.052653$ & $0.0523501$ & $0.0518654$ & $0.0500443$ & $0.0488191$ & $0.050699$ & $0.0684818$ & $0.114494$\\
$1.3693$ & $0.0724616$ & $0.0568817$ & $0.0510125$ & $0.0501911$ & $0.049856$ & $0.0494814$ & $0.0478177$ & $0.0468376$ & $0.0485793$ & $0.0659898$ & $0.111661$\\
$2.28217$ & $0.070031$ & $0.0544353$ & $0.0486847$ & $0.0478525$ & $0.0475057$ & $0.0472524$ & $0.0457502$ & $0.0450143$ & $0.0466329$ & $0.0637098$ & $0.108938$\\
\ldots & \ldots & \ldots & \ldots & \ldots & \ldots & \ldots & \ldots & \ldots & \ldots & \ldots & \ldots \\
$1139.71$ & $0.0795899$ & $0.0646117$ & $0.0585712$ & $0.0578997$ & $0.0577296$ & $0.0570707$ & $0.0549619$ & $0.0532615$ & $0.0554679$ & $0.0741066$ & $0.120519$\\
$1140.63$ & $0.0772525$ & $0.0619816$ & $0.0559683$ & $0.0552269$ & $0.0549786$ & $0.054398$ & $0.0524273$ & $0.0509605$ & $0.0529949$ & $0.0711876$ & $0.117445$\\
$0.456435$ & $0.0156331$ & $0.0160474$ & $0.0163963$ & $0.0163615$ & $0.0158782$ & $0.015466$ & $0.0152769$ & $0.0141906$ & $0.0136551$ & $0.0121923$ & $0.0112008$\\
\ldots & \ldots & \ldots & \ldots & \ldots & \ldots & \ldots & \ldots & \ldots & \ldots & \ldots & \ldots \\\hline
\end{tabular}
\end{table*}

\begin{table*}
 \centering
 \caption{\label{tab:specfidz24} Same as Table~\ref{tab:specfidz2} but for
 the fiducial simulation at $z=2.4$.}
\begin{tabular}{llllllllllll}
\hline\hline &
\multicolumn{11}{c}{$\tau_{\rm HI}$ at transverse separation $\Delta r$ in $h^{-1}$\,kpc comoving}\\
\cline{2-12}\\
$v$ [\kms] & $0$ & $50$ & $100$ & $125$ & $150$ & $175$ & $200$ & $250$ & $300$ & $400$ & $500$\\\hline
$0.478246$ & $0.0199492$ & $0.0187034$ & $0.0163745$ & $0.0154353$ & $0.0144006$ & $0.0132077$ & $0.0122355$ & $0.0116467$ & $0.0139285$ & $0.0237159$ & $0.0437492$\\
$1.43474$ & $0.0201131$ & $0.0187672$ & $0.0163332$ & $0.0153707$ & $0.0143052$ & $0.0130611$ & $0.0120548$ & $0.0114272$ & $0.0136959$ & $0.0237212$ & $0.0449779$\\
$2.39123$ & $0.0202882$ & $0.0188474$ & $0.0163076$ & $0.0153204$ & $0.0142259$ & $0.0129265$ & $0.0118848$ & $0.0112129$ & $0.0134618$ & $0.0237192$ & $0.0462462$\\
\ldots & \ldots & \ldots & \ldots & \ldots & \ldots & \ldots & \ldots & \ldots & \ldots & \ldots & \ldots \\
$1194.18$ & $0.0196609$ & $0.0186309$ & $0.0165093$ & $0.0156127$ & $0.0146395$ & $0.0135399$ & $0.0126293$ & $0.0120978$ & $0.0143827$ & $0.0236818$ & $0.0414031$\\
$1195.14$ & $0.0197978$ & $0.0186573$ & $0.0164328$ & $0.0155157$ & $0.014512$ & $0.0133673$ & $0.0124272$ & $0.0118706$ & $0.0141579$ & $0.0237028$ & $0.0425582$\\
$0.478246$ & $0.192298$ & $0.100375$ & $0.06954$ & $0.0723889$ & $0.0800862$ & $0.0890743$ & $0.100284$ & $0.130356$ & $0.165208$ & $0.133648$ & $0.0857822$\\
\ldots & \ldots & \ldots & \ldots & \ldots & \ldots & \ldots & \ldots & \ldots & \ldots & \ldots & \ldots \\\hline
\end{tabular}
\end{table*}

\begin{table*}
 \centering
 \caption{\label{tab:specH4z24} Same as Table~\ref{tab:specfidz2} but for
 the H4 simulation at $z=2.4$.}
\begin{tabular}{llllllllllll}
\hline\hline &
\multicolumn{11}{c}{$\tau_{\rm HI}$ at transverse separation $\Delta r$ in $h^{-1}$\,kpc comoving}\\
\cline{2-12}\\
$v$ [\kms] & $0$ & $50$ & $100$ & $125$ & $150$ & $175$ & $200$ & $250$ & $300$ & $400$ & $500$\\\hline
$0.478246$ & $0.0163039$ & $0.0154859$ & $0.0148815$ & $0.0151827$ & $0.0158883$ & $0.0167197$ & $0.0179085$ & $0.0193029$ & $0.0189321$ & $0.0209639$ & $0.0399029$\\
$1.43474$ & $0.0162974$ & $0.0154543$ & $0.0147949$ & $0.0151115$ & $0.0158418$ & $0.0167008$ & $0.017951$ & $0.0194545$ & $0.0191134$ & $0.0213779$ & $0.0404602$\\
$2.39123$ & $0.0163207$ & $0.015448$ & $0.0147261$ & $0.0150577$ & $0.0158105$ & $0.0166931$ & $0.0180017$ & $0.01961$ & $0.0193001$ & $0.0218202$ & $0.041028$\\
\ldots & \ldots & \ldots & \ldots & \ldots & \ldots & \ldots & \ldots & \ldots & \ldots & \ldots & \ldots \\
$1194.18$ & $0.0164018$ & $0.0156207$ & $0.015107$ & $0.0153766$ & $0.016026$ & $0.0167901$ & $0.0178472$ & $0.0190116$ & $0.0185841$ & $0.0202135$ & $0.0388243$\\
$1195.14$ & $0.0163391$ & $0.0155417$ & $0.0149856$ & $0.0152711$ & $0.0159498$ & $0.0167495$ & $0.0178739$ & $0.0191552$ & $0.0187557$ & $0.0205764$ & $0.0393573$\\
$0.478246$ & $0.210983$ & $0.230244$ & $0.243747$ & $0.222563$ & $0.192481$ & $0.168656$ & $0.14649$ & $0.12604$ & $0.122942$ & $0.116694$ & $0.134803$\\
\ldots & \ldots & \ldots & \ldots & \ldots & \ldots & \ldots & \ldots & \ldots & \ldots & \ldots & \ldots \\\hline
\end{tabular}
\end{table*}

\begin{table*}
 \centering
 \caption{\label{tab:specfidz3} Same as Table~\ref{tab:specfidz2} but for
 the fiducial simulation at $z=3$.}
\begin{tabular}{llllllllllll}
\hline\hline &
\multicolumn{11}{c}{$\tau_{\rm HI}$ at transverse separation $\Delta r$ in $h^{-1}$\,kpc comoving}\\
\cline{2-12}\\
$v$ [\kms] & $0$ & $50$ & $100$ & $125$ & $150$ & $175$ & $200$ & $250$ & $300$ & $400$ & $500$\\\hline
$0.511584$ & $2.8116$ & $2.65821$ & $1.90155$ & $1.56401$ & $1.246$ & $1.01155$ & $0.779778$ & $0.534622$ & $0.451276$ & $0.441648$ & $0.579892$\\
$1.53475$ & $2.81212$ & $2.6316$ & $1.8672$ & $1.53038$ & $1.21946$ & $0.995949$ & $0.773377$ & $0.534213$ & $0.453597$ & $0.4446$ & $0.577358$\\
$2.55792$ & $2.80893$ & $2.59779$ & $1.82805$ & $1.49217$ & $1.18909$ & $0.97737$ & $0.765317$ & $0.533875$ & $0.456641$ & $0.448473$ & $0.574669$\\
\ldots & \ldots & \ldots & \ldots & \ldots & \ldots & \ldots & \ldots & \ldots & \ldots & \ldots & \ldots \\
$1277.43$ & $2.7989$ & $2.68987$ & $1.9559$ & $1.61738$ & $1.28748$ & $1.03388$ & $0.787756$ & $0.535821$ & $0.449541$ & $0.439144$ & $0.584746$\\
$1278.45$ & $2.80722$ & $2.6776$ & $1.93109$ & $1.59299$ & $1.26864$ & $1.02416$ & $0.784539$ & $0.53515$ & $0.449883$ & $0.439805$ & $0.582338$\\
$0.511584$ & $0.176082$ & $0.167829$ & $0.121278$ & $0.126108$ & $0.10553$ & $0.0909143$ & $0.151623$ & $0.0766237$ & $0.172581$ & $0.299647$ & $0.534714$\\
\ldots & \ldots & \ldots & \ldots & \ldots & \ldots & \ldots & \ldots & \ldots & \ldots & \ldots & \ldots \\\hline
\end{tabular}
\end{table*}

\begin{table*}
 \centering
 \caption{\label{tab:specH4z3} Same as Table~\ref{tab:specfidz2} but for
 the H4 simulation at $z=3$.}
\begin{tabular}{llllllllllll}
\hline\hline &
\multicolumn{11}{c}{$\tau_{\rm HI}$ at transverse separation $\Delta r$ in $h^{-1}$\,kpc comoving}\\
\cline{2-12}\\
$v$ [\kms] & $0$ & $50$ & $100$ & $125$ & $150$ & $175$ & $200$ & $250$ & $300$ & $400$ & $500$\\\hline
$0.511584$ & $1.34765$ & $1.25107$ & $1.10045$ & $1.01987$ & $0.954173$ & $0.851263$ & $0.789765$ & $0.620737$ & $0.456376$ & $0.399912$ & $0.738452$\\
$1.53475$ & $1.37228$ & $1.25437$ & $1.09455$ & $1.00868$ & $0.940173$ & $0.836138$ & $0.773257$ & $0.606903$ & $0.445488$ & $0.396662$ & $0.752027$\\
$2.55792$ & $1.39985$ & $1.25941$ & $1.08958$ & $0.998428$ & $0.926976$ & $0.82169$ & $0.757278$ & $0.593245$ & $0.43461$ & $0.393098$ & $0.765066$\\
\ldots & \ldots & \ldots & \ldots & \ldots & \ldots & \ldots & \ldots & \ldots & \ldots & \ldots & \ldots \\
$1277.43$ & $1.30663$ & $1.24965$ & $1.11538$ & $1.04534$ & $0.984799$ & $0.883627$ & $0.824378$ & $0.648941$ & $0.478036$ & $0.405521$ & $0.710004$\\
$1278.45$ & $1.32581$ & $1.24949$ & $1.10737$ & $1.03206$ & $0.969032$ & $0.867087$ & $0.806807$ & $0.634749$ & $0.467237$ & $0.402857$ & $0.724417$\\
$0.511584$ & $0.23695$ & $0.182967$ & $0.146573$ & $0.153392$ & $0.138004$ & $0.126327$ & $0.177061$ & $0.0988852$ & $0.264086$ & $0.350753$ & $0.399363$\\
\ldots & \ldots & \ldots & \ldots & \ldots & \ldots & \ldots & \ldots & \ldots & \ldots & \ldots & \ldots \\\hline
\end{tabular}
\end{table*}

\begin{table*}
 \centering
 \caption{\label{tab:specfidz4} Same as Table~\ref{tab:specfidz2} but for
 the fiducial simulation at $z=4$.}
\begin{tabular}{llllllllllll}
\hline\hline &
\multicolumn{11}{c}{$\tau_{\rm HI}$ at transverse separation $\Delta r$ in $h^{-1}$\,kpc comoving}\\
\cline{2-12}\\
$v$ [\kms] & $0$ & $50$ & $100$ & $125$ & $150$ & $175$ & $200$ & $250$ & $300$ & $400$ & $500$\\\hline
$0.565685$ & $0.661079$ & $0.505036$ & $0.416128$ & $0.39417$ & $0.393508$ & $0.400554$ & $0.411289$ & $0.438301$ & $0.381488$ & $0.285883$ & $0.235278$\\
$1.69705$ & $0.66984$ & $0.513474$ & $0.42443$ & $0.402405$ & $0.400914$ & $0.406196$ & $0.41516$ & $0.439256$ & $0.36292$ & $0.272959$ & $0.225943$\\
$2.82843$ & $0.678133$ & $0.521662$ & $0.432352$ & $0.410354$ & $0.40801$ & $0.41144$ & $0.418532$ & $0.439755$ & $0.344936$ & $0.260942$ & $0.217593$\\
\ldots & \ldots & \ldots & \ldots & \ldots & \ldots & \ldots & \ldots & \ldots & \ldots & \ldots & \ldots \\
$1412.52$ & $0.642551$ & $0.487853$ & $0.398949$ & $0.377328$ & $0.378122$ & $0.388281$ & $0.40212$ & $0.434716$ & $0.420145$ & $0.314431$ & $0.257342$\\
$1413.65$ & $0.651941$ & $0.496458$ & $0.407588$ & $0.385772$ & $0.385881$ & $0.394565$ & $0.406935$ & $0.436812$ & $0.40058$ & $0.299712$ & $0.245707$\\
$0.565685$ & $1.14591$ & $0.685482$ & $0.457377$ & $0.397335$ & $0.380256$ & $0.329011$ & $0.289736$ & $0.240664$ & $0.196483$ & $0.165188$ & $0.209743$\\
\ldots & \ldots & \ldots & \ldots & \ldots & \ldots & \ldots & \ldots & \ldots & \ldots & \ldots & \ldots \\\hline
\end{tabular}
\end{table*}

\begin{table*}
 \centering
 \caption{\label{tab:specH4z4} Same as Table~\ref{tab:specfidz2} but for
 the H4 simulation at $z=4$.}
\begin{tabular}{llllllllllll}
\hline\hline &
\multicolumn{11}{c}{$\tau_{\rm HI}$ at transverse separation $\Delta r$ in $h^{-1}$\,kpc comoving}\\
\cline{2-12}\\
$v$ [\kms] & $0$ & $50$ & $100$ & $125$ & $150$ & $175$ & $200$ & $250$ & $300$ & $400$ & $500$\\\hline
$0.565685$ & $0.4677$ & $0.455804$ & $0.409949$ & $0.377138$ & $0.347536$ & $0.326676$ & $0.312284$ & $0.301144$ & $0.375485$ & $0.430507$ & $0.451228$\\
$1.69705$ & $0.464439$ & $0.452606$ & $0.406987$ & $0.374329$ & $0.344516$ & $0.323349$ & $0.309021$ & $0.299182$ & $0.364959$ & $0.413616$ & $0.434875$\\
$2.82843$ & $0.461309$ & $0.449434$ & $0.403865$ & $0.371305$ & $0.341251$ & $0.319763$ & $0.305491$ & $0.296945$ & $0.354758$ & $0.39729$ & $0.418832$\\
\ldots & \ldots & \ldots & \ldots & \ldots & \ldots & \ldots & \ldots & \ldots & \ldots & \ldots & \ldots \\
$1412.52$ & $0.474208$ & $0.461919$ & $0.41517$ & $0.381926$ & $0.352658$ & $0.332383$ & $0.317855$ & $0.304162$ & $0.397479$ & $0.466041$ & $0.484875$\\
$1413.65$ & $0.470993$ & $0.458939$ & $0.412695$ & $0.379685$ & $0.350263$ & $0.329701$ & $0.315241$ & $0.30281$ & $0.386327$ & $0.447977$ & $0.467894$\\
$0.565685$ & $1.92588$ & $1.13413$ & $0.655367$ & $0.580232$ & $0.546141$ & $0.521737$ & $0.475962$ & $0.363275$ & $0.298874$ & $0.260041$ & $0.214879$\\
\ldots & \ldots & \ldots & \ldots & \ldots & \ldots & \ldots & \ldots & \ldots & \ldots & \ldots & \ldots \\\hline
\end{tabular}
\end{table*}

\end{landscape}
\end{onecolumn}

\end{document}